\documentclass[aps,prl,twocolumn,groupedaddress,showpacs]{revtex4-1}

\usepackage{graphicx}
\usepackage{amsmath}
\usepackage{subfigure}
\usepackage{amssymb}
\usepackage{amsbsy}
\usepackage{epstopdf}
\usepackage{color}
\usepackage{bm}
\usepackage{setspace}
\usepackage{tikz}
\newcommand*\circled[1]{\tikz[baseline=(char.base)]{
            \node[shape=circle,draw,inner sep=1pt] (char) {#1};}}

\begin{document}

\setlength\fboxsep{0pt}
\setlength\fboxrule{0pt}

\title{Long-range interactions and phase defects in chains of fluid-coupled oscillators}
\author{Douglas R. Brumley$^{1,2}$, Nicolas Bruot$^{3,4}$, Jurij Kotar$^4$, Raymond E. Goldstein$^5$, Pietro Cicuta$^4$ and Marco Polin$^6$}
\email{M.Polin@warwick.ac.uk}

\affiliation{$^1$Ralph M. Parsons Laboratory, Department of Civil and Environmental Engineering, Massachusetts Institute of Technology, Cambridge, MA 02139, USA \\
$^2$Department of Civil, Environmental and Geomatic Engineering, ETH Z\"urich, 8093 Z\"urich, Switzerland \\
$^3$Institute of Industrial Science, University of Tokyo, 4-6-1 Komaba, Meguro-ku, Tokyo 153-8505, Japan \\
$^4$Cavendish Laboratory, University of Cambridge, Cambridge CB3 0HE, UK \\
$^5$Department of Applied Mathematics and Theoretical Physics, University of Cambridge, Centre for Mathematical Sciences, Wilberforce Road, Cambridge CB3 0WA, UK \\
$^6$Physics Department, University of Warwick, Gibbet Hill Road, Coventry CV4 7AL, UK.}

\begin{abstract}
Eukaryotic cilia and flagella are chemo-mechanical oscillators capable of generating long-range coordinated motions known as metachronal waves. Pair synchronization is a fundamental requirement for these collective dynamics, but it is generally not sufficient for collective phase-locking, chiefly due to the effect of long-range interactions. Here we explore experimentally and numerically a minimal model for a ciliated surface; hydrodynamically coupled oscillators rotating above a no-slip plane. Increasing their distance from the wall profoundly effects the global dynamics, due to variations in hydrodynamic interaction range. The array undergoes a transition from a traveling wave to either a steady chevron pattern or one punctuated by periodic phase defects. Within the transition between these regimes the system displays behavior reminiscent of chimera states. 
\end{abstract}

\maketitle

The ability of ensembles of oscillators to achieve 
collective motions is fundamental in biological processes ranging
from the initiation of heartbeats to the motility of microorganisms. The emergent properties of coupled oscillators 
can vary dramatically depending on the intrinsic properties of the oscillators and the nature of the coupling between 
them \cite{Dorfler2014}. Flashing fireflies equally and instantaneously coupled to one another \cite{Mirollo1990} can 
support very different behaviors to chemical micro-oscillators which are coupled only locally, and subject to time delays \cite{Toiya2010}.

Eukaryotic cilia and flagella are chemo-mechanical oscillators which generate a variety of collective 
motions that can be quantified with high-speed microscopy in microfluidic environments 
\cite{Son2015,Goldstein2015,Quaranta2016}. The molecular biology of these internally driven filaments is virtually identical 
in green algae \cite{Goldstein2015}, protists \cite{Sleigh1962} and humans \cite{Button:2012}, and the flows they generate 
fulfill crucial roles in development, motility, sensing and transport. When close together, the mutual interaction 
between their oscillatory flow fields can cause them to beat in synchrony \cite{Brumley2014}, and larger ensembles of flagella 
demonstrate striking collective motions in the form of metachronal waves (MWs) \cite{Knight_Jones:1954,Brumley:2012,Brumley:2015_JRSI}, 
akin to the `Mexican wave' propagating around a packed stadium. Many surrogate models for flagellar dynamics have been 
proposed \cite{Brumley:2015_JRSI, Niedermayer:2008fk, Vilfan:2006uq, Uchida:2011kx, Wollin:2011, Gueron:1999, Lagomarsino03, Osterman:2011,
Elgeti:2013, Bruot13, Kavre2015}, typically with a set geometry which fixes the range and coupling between oscillators.

Here we relax this condition and study a linear array of colloidal oscillators \cite{Bruot2016} driven in circular trajectories 
at a controllable height above a no-slip wall.
The oscillator couplings can be modified continuously from being primarily through nearest neighbors
to a regime involving significant long-range interactions. As a function of rotor properties, 
a traveling wave found at small heights becomes either a chevron pattern 
or is punctuated by phase defects at large ones. 
The transition is not a gradual morphing between the two profiles, but rather a process involving generation and 
propagation of defects along the strip, 
where phase-locked and non-phase-locked subgroups of oscillators can coexist. A behavior arising from 
long-range interactions whose amplitude is modulated by the distance from the wall \cite{Wollin:2011}, these dynamics are 
reminiscent of chimera states, in which oscillators split into phase-locked and desynchronized clusters \cite{Abrams2004,Martens2013}.

\begin{figure}[h!]
\begin{center}
\includegraphics[width=\columnwidth]{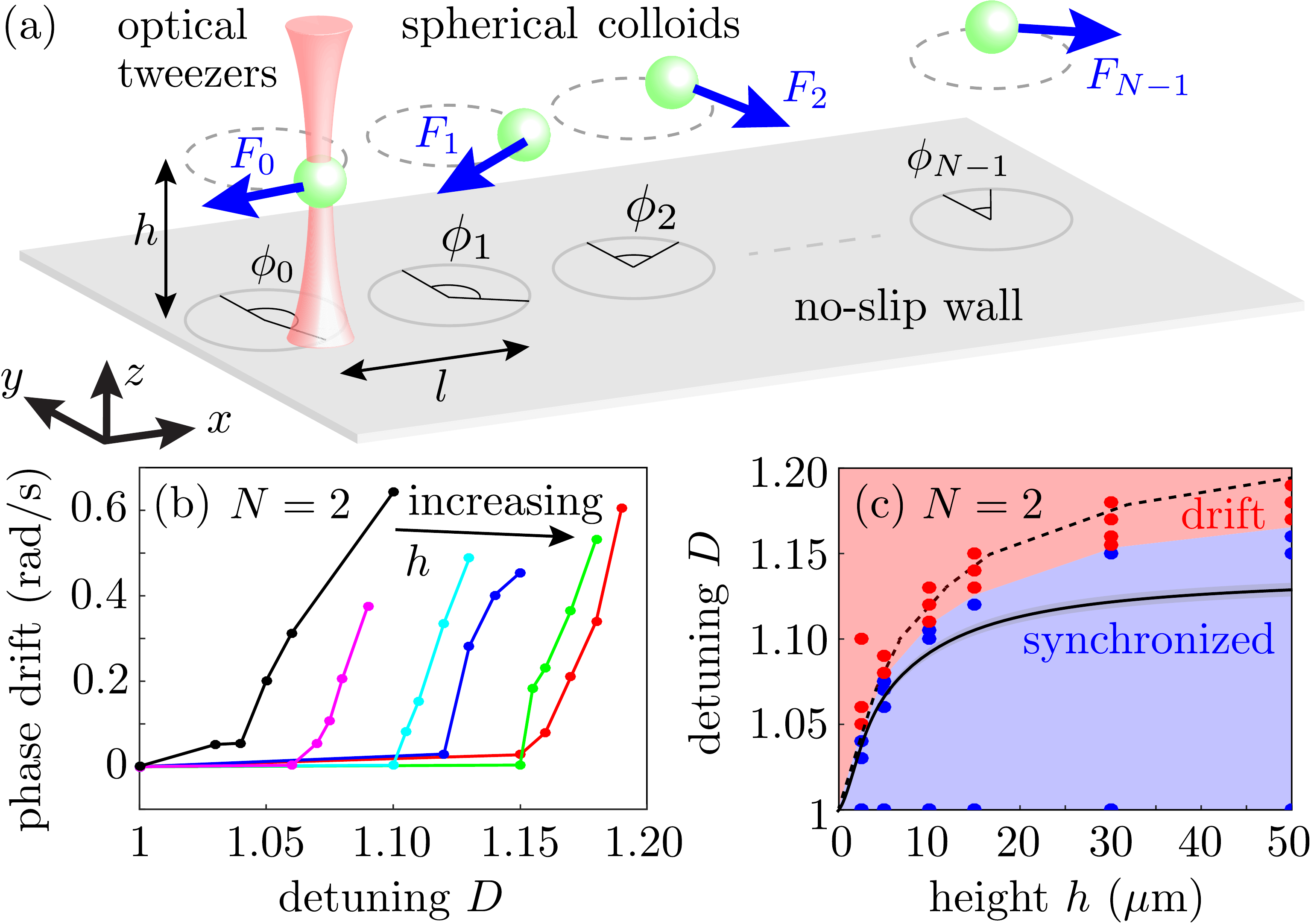} 
\caption{(color online). Experimental setup and results.  (a) Microspheres a distance $h$ above a no-slip boundary
are driven in circular trajectories by time-sharing optical tweezers. (b) Average phase drift $\chi = \phi_1 - \phi_0$ for a rotor pair vs. detuning $D$ for 
$h=4.2({\color{black}\bullet}), 6.7 ({\color{magenta}\bullet}), 11.7 ({\color{cyan}\bullet}), 16.7 ({\color{blue}\bullet}), 
31.7 ({\color{green}\bullet}), 51.7 ({\color{red}\bullet})\,\mu$m. (c) Phase diagram showing experimental regions of 
synchrony (blue) and drift (red), the boundary from hydrodynamic simulations (dashed) and theory from Eq.~\eqref{two_rot_avg_drift} (solid).}
\label{fig1}
\end{center}
\end{figure}

In our experiments, silica colloids of radius $a=1.74\,\mu$m (BangsLab, USA) suspended in a water-glycerol solution of 
viscosity $\mu=6\,$mPas within a $150\, \mu$m-thick sample, are captured and driven by feedback-controlled time-shared 
(20~kHz) optical tweezers (OTs) based on acousto-optical deflection of a $1064~$nm-wavelength diode-pumped solid-state laser 
(CrystaLaser IRCL-2W-1064) as previously described \cite{Leoni2009,Kotar2013}. The OTs describe a planar array of circular trajectories (Fig.~\ref{fig1}a) of radius $R=1.59\,\mu$m and center-to-center separation $\ell=9.19\,\mu$m, a distance $h$ above the sample bottom, with $4.2(4)\, \mu$m $\le h \le 51.7(4)\,\mu$m.
This configuration, which reflects the capabilities and limitations of our OT setup, is similar to arrays of nodal cilia, but differs 
from the more common situation where the ciliary beating plane is perpendicular to an organism's surface.
The OTs are arranged so that a colloidal particle on the $i$th trajectory ($i\in\{0,\ldots,N-1\}$) experiences a 
radial harmonic potential with spring constant $\lambda = 2.06\pm0.06\,$pN/$\mu$m resisting excursions from the prescribed radius, 
and a constant tangential force of magnitude $F_i = F_{\text{dr}}D^{i-0.5}$ leading to rotation. Before each experiment we 
calibrate $F_{\text{dr}}\simeq 2.23\,$pN (see \cite{SM}; typical variation $\pm 2\%$). $D\neq 1$ is used to break left-right symmetry 
along the chain and induce a stable traveling wave for small $h$ \cite{Brumley:2012}.
For the detuning adopted here, $D=1.01$, the period of individual oscillators varies between $\tau\sim0.5\,$s and $\sim1\,$s across the explored range of $h$. 
The oscillators are imaged under a Nikon inverted Eclipse Ti-E with a $60\times$ Nikon Plan Apo VC water immersion objective ($\text{NA}=1.20$), and 
recorded for up to $1200\,$s using an AVT Marlin F131B CMOS camera set at $229\,$fps. The particle's positions are measured using a 
profile image correlation with subpixel resolution, and used to track the rotor phases $\{\phi_i(t)\}$ (Fig.~\ref{fig1}a).  
Isolated oscillators rotate with a height-dependent angular velocity $\omega_i = F_i/R\zeta_0\zeta_w$, where $\zeta_0 = 6\pi \mu a$ is the sphere's 
bulk drag coefficient, and $\zeta_w(h) = 1+\frac{9}{16}\frac{a}{h} + \mathcal{O}(a^3/h^3)$ accounts for the presence of the wall \cite{Happel_Brenner}.
Experimental results are compared with deterministic hydrodynamic simulations based on the Oseen approximation 
to inter-particle coupling \cite{Brumley:2012}. Unavoidable delays in the OT's feedback response introduce a mismatch 
between experiments and simulations which for a pair of oscillators is corrected by increasing the simulation value of $\lambda$ 
by a factor $\kappa$ relative to the experimental one. We estimate $\kappa = 2.21$ in agreement with previous reports \cite{Kotar2013}. 

Consider first two rotors separated by a distance $\ell$. 
For rotors with instantaneous positions $\{\bm{x}_i\}$ and velocities $\{\bm{v}_i\}$, the hydrodynamic drag on the 
$i$th rotor is given by $-\bm{\zeta}(\bm{x}_i) \cdot [\bm{v}_i - \sum_{j \neq i} \mathbf{G}(\bm{x}_j,\bm{x}_i) \cdot \bm{F}_j^{\text{ext}}]$, 
where $\bm{F}_j^{\text{ext}}$ is the net external force acting on the $j$th sphere and 
$\mathbf{G}(\bm{x}_i,\bm{x}_j)$ is the Green's function in the presence of the no-slip wall. For identical 
rotors (detuning $D=1$), hydrodynamic coupling eventually leads to synchrony provided $\lambda < \infty$, by perturbing the angular 
velocities of the two rotors so that the leading and lagging rotors become slower and faster respectively \cite{Niedermayer:2008fk, Brumley:2015_JRSI}. 
The timescale for synchronization depends on the spring constant $\lambda$ and the strength of hydrodynamic interactions between rotors, 
which is itself a function of height $h$ and spacing $\ell$. The dynamics become richer if a discrepancy between the rotor's intrinsic frequencies 
is introduced ($D \neq 1$), for then the coupling must be sufficiently strong to overcome the natural tendency for the rotor's 
phase difference $\chi = \phi_1 - \phi_0$ to drift.

\begin{figure*}[t!]
\begin{center}
\includegraphics[width=\textwidth]{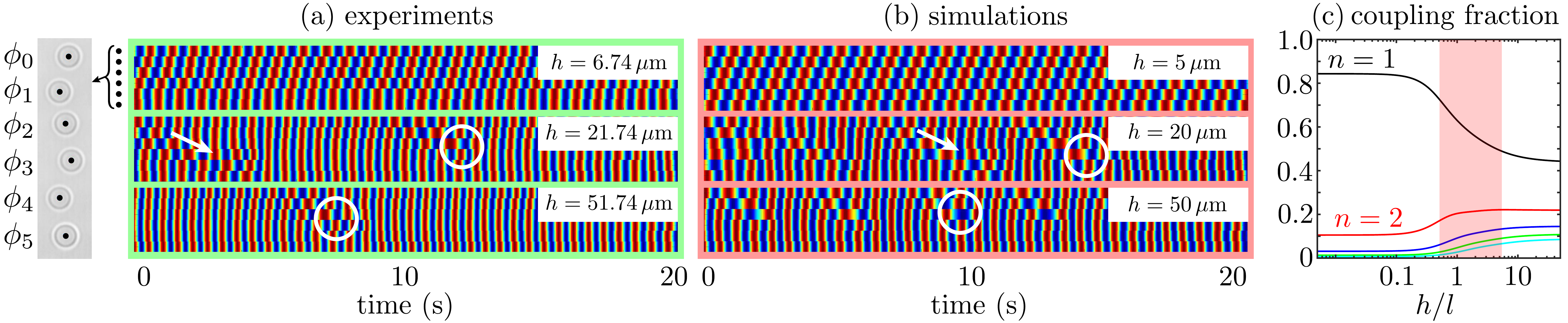} 
\caption{(color online). Results for the linear array of driven colloidal oscillators. (a) Kymographs showing $\sin \phi_i$ at three 
heights above the wall. With increasing $h$, the traveling wave becomes frustrated, with the introduction of wobbles (arrows) and phase defects (circles). 
(b) Numerical results from model. (c) Fraction of total coupling corresponding to interacting with different
neighbors, as a function of $h$. Shaded red region represents experimental parameter regime.}
\label{fig2}
\end{center}
\end{figure*}

Bifurcation plots in Fig.~\ref{fig1}b show, for different $h$, the average phase drift between two oscillators as a function of $D$. 
The behavior is typical of a saddle-node bifurcation: the oscillators phase-lock until $D$ reaches a critical  
value $D^*(h)$ and then drift with a monotonically increasing speed. $D^*(h)$ increases with $h$, reflecting the strengthening of 
inter-rotor hydrodynamic coupling with increasing distance from the wall. The phase-locking behavior is summarized
in Fig.~\ref{fig1}c, where the experimental synchronization boundary is based on a threshold of 5 slips in the whole 
experiment ($\dot{\chi}_{\text{av}} = 0.131\,$rad/s). As $h$ is increased, the rotor pair moves deeper into the synchronized region: 
the coupling between the two strengthens due to lower hydrodynamic screening from the wall, leading to an enhanced stability of the synchronized state.
This is reproduced by simulations (Fig.~\ref{fig1}c) up to small shift in $D$, likely the result of a finite value of 
$a/d$ and experimental noise. In the limit $a, R\ll\ell$, the evolution of the phase difference $\chi = \phi_1-\phi_0$ can be 
derived by a generalization of previous arguments \cite{Niedermayer:2008fk,SM}. As phase-locking is slow compared to the  
rotation period, we average over 
this fast timescale and find
\begin{equation}
\dot{\chi} = \frac{F_1-F_0}{R_0 \zeta_0\zeta_w} - \frac{3a}{4\ell}\frac{F_0 F_1}{\lambda \zeta_0 R_0^2}\big[ 2A(\beta)+B(\beta) \big] \sin \chi,
\label{two_rot_chi}
\end{equation}
where, $A(\beta)=1-X-\tfrac{\beta^2}{2}X^{3}$, $B(\beta)=1-X^{3}+\tfrac{3\beta^2}{2}X^{5}$, $X = 1/\sqrt{1+\beta^2}$, and $\beta = 2h/\ell$. From Eq.~\eqref{two_rot_chi}, the average phase drift $\dot{\chi}_{\text{av}}$ for non-phase-locked states reads
\begin{equation}
\dot{\chi}_{\text{av}} = \sqrt{\left(\frac{F_1-F_0}{R_0\zeta_0\zeta_w} \right)^2 - \left(\frac{3a}{4\ell}\frac{F_0 F_1}{\lambda\zeta_0R_0^2}\left(2A(\beta)+B(\beta)\right)\right)^2}. \label{two_rot_avg_drift}
\end{equation}
Given the functional form of the frequency detuning, $F_i = F_{\text{dr}} D^{i-0.5}$, Eq.~\eqref{two_rot_avg_drift} can be solved explicitly to yield the critical detuning $D^*(h)$ (solid line in Fig.~\ref{fig1}c).
The theoretical and numerical solutions for the boundary in Fig.~\ref{fig1}c slightly under- and over-estimate the data, 
respectively, owing to neglect of temporal variations in the inter-particle spacing and the finite size of the beads, respectively. 
Both also neglect thermal fluctuations.

We now turn to the dynamics of a linear array of 6 rotors, with the $i$th rotor centered at $\bm{x} = (i l,0,h)$. 
This is the longest controllable chain with our active-feedback-based OTs. 
The dynamics are studied experimentally as a function of $h$, 
but numerical simulations allow wider exploration of parameters, including changes in the radial stiffness $\lambda$, which 
governs the coupling strength \cite{Niedermayer:2008fk,Brumley:2012,Kotar2013,Brumley2014,Brumley:2015_JRSI} as in Eq.~\eqref{two_rot_chi}. 
In both experiments and simulations we introduce a mild frequency bias $D=1.01$, typical also of {\it Volvox} colonies \cite{Brumley:2015_JRSI}, 
which breaks the translational symmetry and induces a MW for $h\lesssim 10\, \mu$m. At all heights studied, this value 
of $D$ is deep within the synchronized region of parameter space for two rotors.

\begin{figure}[htp!]
\begin{center}
\includegraphics[width=\columnwidth]{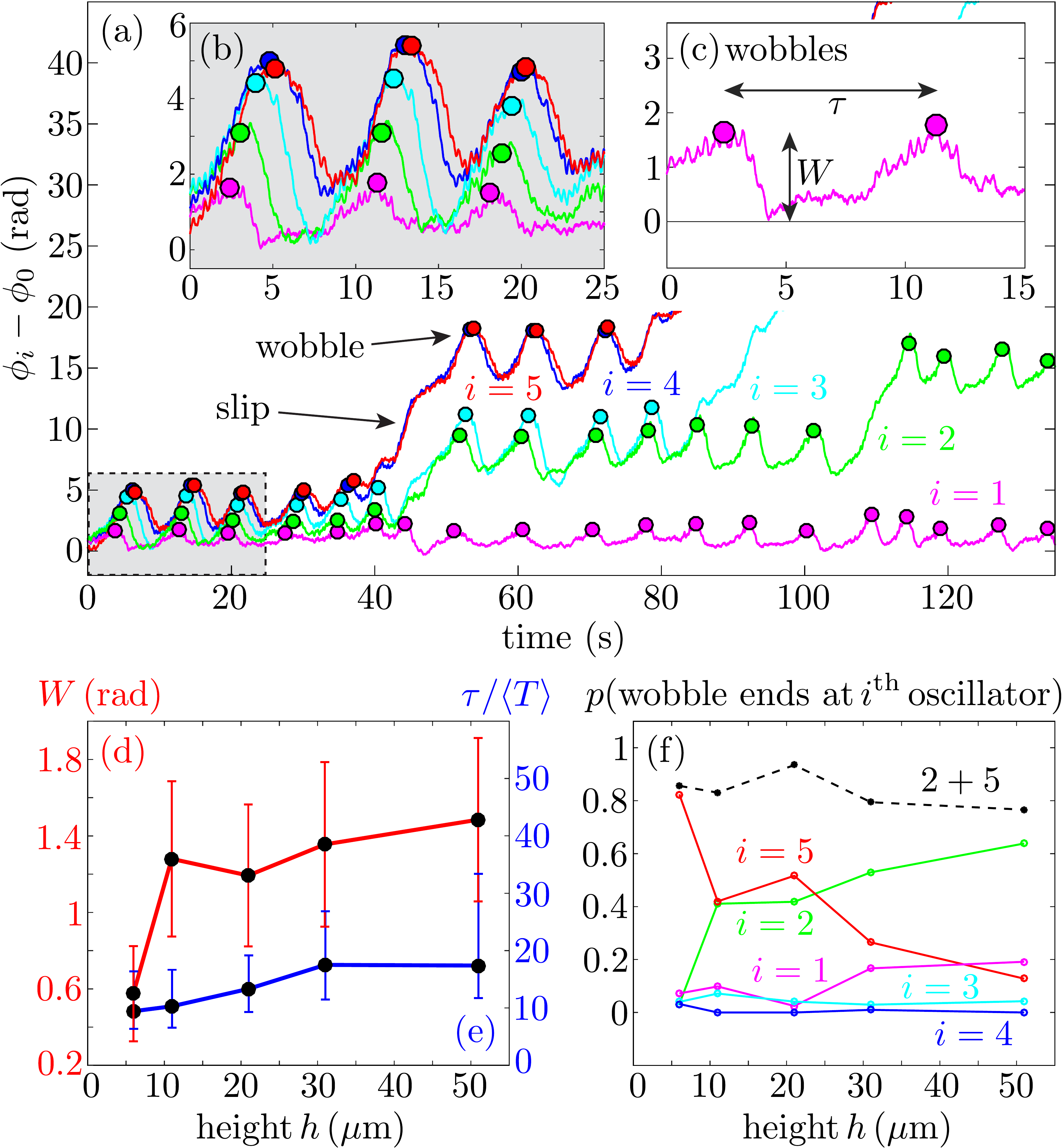} 
\caption{(color online). Experimental phase dynamics. (a,b) Phase difference relative to the first rotor, $\phi_i-\phi_0$, at $h=11.7\, \mu$m. 
(c) Wobbles are characterized by their magnitude $W$ (radians) and timescale $\tau / \langle T \rangle$ (normalized by rotor period), shown as a 
function of $h$ in panels (d) and (e). (f) Probability that a propagating wobble ends at rotor $i$, resulting in a slip.}
\label{fig3}
\end{center}
\end{figure}

\begin{figure*}[htp!]
\begin{center}
\includegraphics[width=\textwidth]{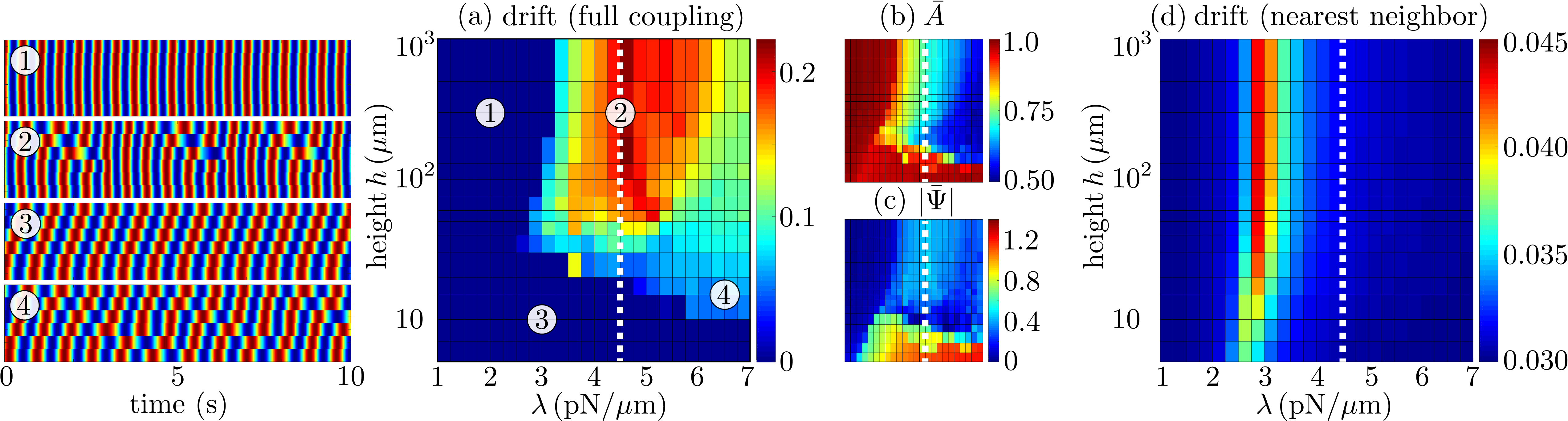} 
\caption{(color online). (a) Average phase drift per beat between end oscillators (measured in beats) as a function of height above the wall and radial spring stiffness. Shown also are four representative kymographs. (b) Time-averaged amplitude $\bar{A}$ and (c) angle $|\bar{\Psi}|$ of the complex order parameter $Z = A e^{i \Psi}$. The axes are the same as in a. (d) Same as a but with hydrodynamic interactions truncated to nearest neighbor. Parameters used include $a = 1.74 \, \mu$m, $\ell=9.19\,\mu$m, $R=1.59\,\mu$m, viscosity $\mu=6\,$mPas. Simulations correspond to $0 \leq t \leq 2000$~s.  The dashed white line shows the value of $\lambda$ corresponding to Fig.~\ref{fig2}.}
\label{fig4}
\end{center}
\end{figure*}

Figure~\ref{fig2}a shows that at $h=6.7(4)\,\mu$m the rotors phase-lock in a stable MW whose direction is set by the frequency bias. 
With increasing $h$, defects (phase slips) emerge, giving rise to a net drift in the cumulative phase difference between rotors at 
opposite ends of the chain. Phase defects always propagate in the direction of the fastest oscillator. At these intermediate heights, the 
phase profile also displays ``wobbles'' -- perturbations to the MW that are not accompanied by a phase defect. 
Numerical results shown in Fig.~\ref{fig2}b capture the traveling wave at 
$h=5 \, \mu$m, the presence of defects and their propagation direction, and wobbles at larger heights.
At the largest height, $h=50 \, \mu$m, defects no longer propagate through the chain, and rotors 3-5 remain phase-locked.

The phase dynamics of wobbles and defects are shown in Fig.~\ref{fig3}a for $h=11.7(4)\,\mu$m. The first 25 seconds of the time-series show fluctuations 
in $\phi_i-\phi_0$ (wobbles), even while the system is frequency-locked. Fluctuations start at the first oscillator pair and travel unidirectionally 
along the chain (Fig.~\ref{fig3}b) with a preserved, soliton-like signature \cite{Ahnert2008}. Occasionally, they terminate within the chain with a 
slip (Fig.~\ref{fig3}a); these are the phase defects observed in kymographs. 
Both wobbles and defects are characterized by initial excursions of amplitude $W$ and recurrence time $\tau$ (Fig.~\ref{fig3}c) which 
depend on $h$ (Fig.~\ref{fig3}d,e). 
The typical time $\tau \sim 10\,\langle T \rangle$ (where $\langle T \rangle\simeq 1\,$s is the average period) depends less strongly on $h$ 
than does $W$, which shows a pronounced growth (Fig.~\ref{fig3}d) mirroring the increased probability that a wobble will terminate 
in a slip within the chain, causing a defect (Fig.~\ref{fig3}f).
Although their position can vary, defects tend to cluster, in this case at the middle of the chain (position $i=2$), as seen also 
in simulations of longer chains \cite{SM}.

The hydrodynamic coupling between two rotors increases monotonically with $h$; for an isolated pair, this manifests in more robust synchronization at larger heights. 
For a chain of rotors, increasing $h$ has the reverse effect, disrupting the stable MW with wobbles and punctuating it with periodic phase defects (Fig.~\ref{fig2}).  
The hydrodynamic coupling between every pair of rotors in the chain grows as $h$ is increased. For just two rotors, Eq.~\eqref{two_rot_chi} shows that 
equivalent changes to the hydrodynamic coupling can be achieved through modification of the mean interparticle separation $l$. For the chain of 6 rotors, in which longer 
range hydrodynamic interactions also occur, changes to $h$ and $l$ are no longer equivalent.

The peculiar dynamics observed arise from a change in the relative contributions of interactions with different neighbors. The no-slip wall 
has the effect of screening the hydrodynamic interactions in a way that qualitatively changes as a function of $\beta = 2h/\ell$.
This is an important determinant of MW stability, as observed also in simulations of colloidal ``rowers'' \cite{Wollin:2011}.
Figure~\ref{fig2}c shows the magnitude of the coupling of a given oscillator with its $j$th nearest neighbor, estimated with 
Eq.~\eqref{two_rot_chi}, normalized by the total interaction strength with the first 5 neighbors. Although all pairwise couplings 
grow monotonically with $h$, the relative magnitude of the nearest neighbor interactions actually diminishes. Conversely, the relative 
importance of all others increases with $h$. Hydrodynamic disturbances parallel to the wall decay as $u \sim r^{-j}$ where 
$j=1$ and 3 represent the far ($\beta \gg 1$) and near ($\beta \ll 1$) asymptotic limits \cite{Wollin:2011}.  For the end rotor 
the magnitude of the coupling with the $n$th nearest neighbor, normalized by the total coupling strength is
$S(n) = n^{-j} / \sum_{i=1}^{5} i^{-j}$.  For $\beta \ll 1$ the 
interactions are dominated by nearest neighbor, with $S(1)=0.84$, while for $\beta \gg 1$, $S(1) = 0.44$ (see black curve in Fig.~\ref{fig2}c). 
We test the hypothesis that the breakdown of the traveling wave is due to
long-range hydrodynamic interactions through simulations in which interactions are truncated at nearest neighbors, and find the abundance of 
defects is significantly reduced. Importantly, the dynamics are nearly insensitive to $h$, with a maximum 
relative variation in end-to-end drift speed of just $3\%$
between $h=5\,\mu$m and $1000\,\mu$m (Fig.~\ref{fig4}) \cite{SM}.

Additional numerical simulations permit the wider exploration of parameter space. Figure~\ref{fig4}a shows the 
average end-to-end phase drift per beat as a function of $\lambda$ and $h$. We also compute the complex order parameter 
$Z = A e^{i \Psi} = \frac{1}{N-1} \sum_{n=0}^{N-1} e^{i \chi_n}$ where $\chi_n = \phi_{n+1}-\phi_{n}$ \cite{Pikovsky:2003, Wollin:2011}. 
Using the average values $\bar{A}$ and $\bar{\Psi}$ for $t>200$~s [Fig.~\ref{fig4}b,c], and
following the experimental path (dotted line), we see the stable traveling wave at small $h$ $\circled{3}$ developing and, as $h$ increases, 
defects and wobbles along the whole chain first $\circled{4}$ and then localized to the initial half of the chain $\circled{2}$, with 
the remaining three oscillators constantly phase-locked. 
At values of $\lambda$ smaller than the experimental one, however, we observe richer dynamics. For $\lambda \lesssim 2.5\,$pN$/\mu$m, 
the pattern morphs continuously between different types of complete synchronization, going from a MW $\circled{3}$ to 
a chevron-like pattern $\circled{1}$.
These transitions happen without the emergence of defects \cite{Brumley:2012}. 
For $2.5\,\textrm{pN}/\mu\textrm{m} < \lambda < 3\,\textrm{pN}/\mu\textrm{m}$ the system shows reentrant behavior with defects 
only at intermediate heights, separating a MW region from a chevron-like region. The order parameter angle $|\bar{\Psi}|$ (Fig.~\ref{fig4}c) 
identifies clearly the stable MW (yellow/orange) and chevron (dark blue) regions of parameter space. For a fixed $h \gtrsim 50\,\mu$m, 
increasing $\lambda$ results in a monotonic decrease in $\bar{A}$ owing to the reduced rotor compliance. Conversely, the end-to-end phase drift 
exhibits a strong peak around $\lambda = 4.5\,\textrm{pN}/\mu\textrm{m}$, where the rotors slip approximately one beat in every five, despite an intrinsic frequency 
difference of just 5\%. These nontrivial dynamics emerge due to the combination of phase slips induced by long-range interactions, and 
rapid healing of phase defects through orbit compliance. The complete absence of these features from the simulations with nearest neighbor 
coupling alone (Fig.~\ref{fig4}d) highlights the role played by competition between interactions at different ranges. Changing $h$ is then 
a simple and accessible way to modulate their relative strength (see Fig.~\ref{fig2}).

Large arrays of cilia are synonymous with no-slip boundaries, and in many cases, the spacing between these organelles is comparable to their 
length \cite{Brumley:2015_JRSI}, so that effectively $h/\ell \sim 1$ (see Fig.~\ref{fig4}a). 
Our results suggest that flagella of {\it Volvox} may then be balancing the need to extend out into the fluid enough to generate a 
vigorous thrust, with the screening of long-rage hydrodynamic interactions  necessary to stabilize MWs on the colony surface.
As a result, ensembles of flagella in {\it Volvox} \cite{Brumley:2012} (but see also numerical simulations \cite{Elgeti:2013}) may operate in a regime 
naturally prone to the emergence of metachronal phase defects, which are indeed observed experimentally \cite{Brumley:2015_JRSI}.

\begin{acknowledgements}
We are grateful to T.J.~Pedley and D. Bartolo for useful discussions. This work was supported by a Human Frontier Science Program Cross-Disciplinary Fellowship (DRB), a Wellcome Trust Senior Investigator Award (REG), and the EU ERC CoG Hydrosync (PC).
\end{acknowledgements}

\section{Supplementary Material}

\setcounter{figure}{0}
\renewcommand{\thefigure}{S\arabic{figure}}
\setcounter{equation}{0}
\renewcommand{\theequation}{S\arabic{equation}}

\subsection{Single rotor force calibration}

\begin{figure}[thp!]
\begin{center}
\includegraphics[width=0.9\columnwidth]{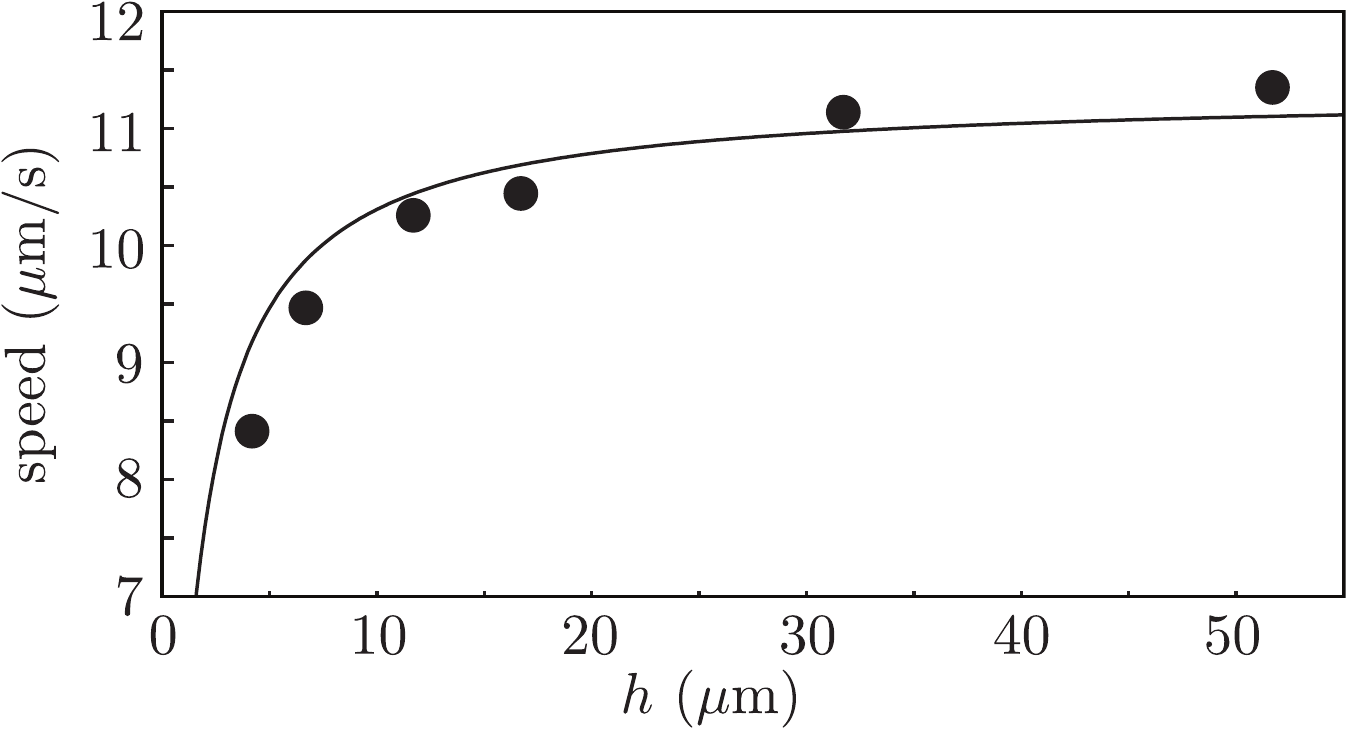} 
\caption{Calibration of an individual colloidal oscillator, moving under the influence of a harmonic potential in an optical tweezer. Experimental results (dots) are shown alongside the prediction of Eq.~\eqref{vel_calibration}, with which the driving force of $F_{\text{dr}} = 2.23~$pN can be extracted.}
\label{1r_calibration}
\end{center}
\end{figure}

For a single bead of radius $a$ in a viscous fluid, situated at a distance $h$ from an infinite no-slip boundary, the external applied force $\bm{F}$ is related to its velocity $\bm{v}$ according to $\bm{F} = \bm{\zeta} \cdot \bm{v}$, where $\bm{\zeta}$ is the anisotropic drag matrix, given by \cite{Vilfan:2006uq}
\begin{equation}
\bm{\zeta} = \bm{\zeta}(h) = \zeta_0 \big[ \mathbf{I} + \tfrac{9a}{16h}(\mathbf{I}+\bm{e}_z \bm{e}_z) + \mathcal{O}((a/h)^3) \big].
\end{equation}
The coefficient $\zeta_0 = 6 \pi \mu a$ is the drag on the sphere in an unbounded fluid of viscosity $\mu$ (equivalent to setting $h \rightarrow \infty$). We are interested only in trajectories which are parallel to the no-slip wall ($\bm{v} \cdot \bm{e}_z =0$). For a constant applied driving force $F_{\text{dr}}$, the sphere's speed $v= |\bm{v}|$ is given by
\begin{equation}
v \simeq \frac{F_{\text{dr}}}{\zeta_0 (1 + \tfrac{9}{16} \tfrac{a}{h})}, 
\label{vel_calibration}
\end{equation}
implying a monotonic increase of the sphere's speed with $h$ for a given $F_{\text{dr}}$. Each set of experiments involves studying the colloidal oscillators at a number of different heights $h$. For each set, the center of the trajectory, its radius and the driving and radial forces are calibrated, for each individually loaded rotor, at the height of $h=22\,\mu$m. These are then checked for independence on $h$. Figure~\ref{1r_calibration} shows, after a full calibration, the speed of an individually loaded colloidal oscillator at 6 different heights together with the prediction from Eq.~\eqref{vel_calibration} using a constant driving force. The two agree well for $F_{\text{dr}} = 2.23\,$pN.

\subsection{Hydrodynamic interaction of two rotors at an arbitrary distance from a no-slip plane}
The fluid disturbance produced by the motion of a sphere parallel to a no-slip wall depends on its height above the planar boundary. For two such spheres situated in the fluid, it is important to calculate the strength of the hydrodynamic interactions between them, and the subsequent effects on their dynamics. We consider two spheres of radius $a$ driven around circular orbits of radius $R_0$ which are parallel to a no-slip wall. The orbit's centers are located at positions $(x,y,z) = (0,0,h)$ and $(\ell,0,h)$ respectively. The plane $z=0$ represents the no-slip boundary, with the semi-infinite domain $z>0$ filled with fluid of viscosity $\mu$. $(\hat{e}_{\phi1},\hat{e}_{R1})$ and $(\hat{e}_{\phi2},\hat{e}_{R2})$ are the unit vectors of the local cylindrical frame of reference of each single rotor. This reference frame is centered at the center of the orbit. The displacement of each sphere from the center of its trajectory will be expressed in its cylindrical frame of reference as $(R_1,\phi_1)$ and $(R_2,\phi_2)$ respectively. Each rotor is subject to a constant tangential driving force $\bm{F}_i = F_i \hat{e}_{Ri}$, and also to a radial spring force with stiffness $\lambda$, which suppresses excursions from the equilibrium radius $R_0$. The spring stiffness is assumed to be large enough that the radial degree of freedom is slaved to the angular degree of freedom. That is, knowing $(\phi_1, \phi_2)$, we know the instantaneous value of $(R_1,R_2)$. It will be our goal to derive the equations of motion of the spheres, without making any assumptions about the relative magnitudes of $\ell$ and $h$.

We assume that sphere radius and trajectory radius are both small compared to other length scales ($a,R_0 \ll h,\ell$). Correspondingly, the two orbits are sufficiently far from each other that we can neglect the variation in the relative separation between the spheres as they move. The separation vector will always be taken to be $\ell \hat{e}_x$. We will write the relation between the sphere's angular velocity $\omega_i$ and the tangential driving force $F_i$, as $F_i = \zeta_0\zeta_w R_0 \omega_i$, where
$\zeta_0 = 6\pi\mu a$ is the bulk drag coefficient, and $\zeta_w$ is the correction due to the presence of the wall. For sufficiently small $a/h$ this correction can be written as $\zeta_w = 1+(9a/16h) + \mathcal{O} (\left(a/h\right)^3 )$. The first thing required is the generic expression of the `Blakelet', i.e. the Stokeslet on top of a bounding wall. This is given by \cite{Blake:1971uq}:
\begin{align}
& v_{21,i} = \frac{F_j}{8\pi\mu} \bigg[ \bigg( \frac{\delta_{ij}}{r} +\frac{r_ir_j}{r^3} \bigg) - \bigg( \frac{\delta_{ij}}{R} +\frac{R_iR_j}{R^3} \bigg) \nonumber \\
 &+2h(\delta_{j\alpha}\delta_{\alpha k} - \delta_{j3}\delta_{3 k})\frac{\partial}{\partial R_k}\left\{\frac{hR_i}{R^3} - \bigg(\frac{\delta_{i3}}{R} +\frac{R_iR_3}{R^3} \bigg)\right\} \bigg]. \label{Blakelet}
\end{align}
Here $\mathbf{r} = (\ell,0,0)$, $r=|\mathbf{r}|$; $\mathbf{R} = (\ell,0,2h)$, $R=|\mathbf{R}|$; $\alpha\in\{1,2\}$. In our case, the point force will be in the $(x,y)$ plane. Furthermore, we are only interested in the component of the velocity in the same plane, since the trajectories are constrained to lie at $z=h$. The background flow is due to the motion of the other sphere, which is:
\begin{align}
\bm{v}_{21,||} &= \frac{1}{8\pi\mu} \bigg[ \bigg(\frac{\mathbf{F}}{r} + \frac{\mathbf{r}(\mathbf{F\cdot r})}{r^3} \bigg) - \bigg(\frac{\mathbf{F}}{R} + \frac{\mathbf{R}(\mathbf{F\cdot R})}{R^3} \bigg)_{||} \nonumber \\
& -2h^2 \bigg( \frac{\mathbf{F}}{R^3} -3\frac{\mathbf{R}(\mathbf{F\cdot R})}{R^5}  \bigg)_{||} \bigg].
\end{align}
In the limit as $h \rightarrow 0$, this reduces to Eq.~(16) in \cite{Niedermayer:2008fk}. Being interested only in the $||$ component, and because $\mathbf{F} = \mathbf{F}_{||}$, we can substitute
\begin{equation}
\mathbf{R}(\mathbf{F\cdot R}) = (\mathbf{r} + 2h\hat{e}_3)(\mathbf{F\cdot r}) \to \mathbf{r}(\mathbf{F\cdot r}).
\end{equation}
Calling, as in \cite{Niedermayer:2008fk}, $\mathbf{s} = \mathbf{F}/8\pi\mu$ we get 
\begin{equation}
\bm{v}_{21,||} = \frac{A(\beta)\mathbf{s} + B(\beta)\hat{r}(\mathbf{s}\cdot\hat{r})}{r}
\end{equation}
where $\hat{r}=\mathbf{r}/r = \hat{e}_x$, called $\mathbf{n}_{21}$ in \cite{Niedermayer:2008fk}; $\beta=2h/\ell$, and 
\begin{align}
A(\beta) &=1-\left(\frac{1}{1+\beta^2}\right)^{\frac{1}{2}}-\frac{\beta^2}{2}\left(\frac{1}{1+\beta^2}\right)^{\frac{3}{2}},
\end{align}
\begin{align}
B(\beta) &=1-\left(\frac{1}{1+\beta^2}\right)^{\frac{3}{2}}+\frac{3\beta^2}{2}\left(\frac{1}{1+\beta^2}\right)^{\frac{5}{2}}.
\end{align}
Notice that, due to the nearby wall, the strength of the Stokeslet $\mathbf{s}$ is written in terms of the sphere's velocity as: 
\begin{equation}
\mathbf{s} = \frac{3}{4}a\zeta_0\zeta_w R_i\dot{\phi}_i\hat{e}_{\phi_i}.
\end{equation}
The derivation of the equations of motion follows the same procedure outlined in Appendix A of \cite{Niedermayer:2008fk}. The only things we need to calculate are:
\begin{align}
\hat{e}_{\phi 1} \cdot \bm{v}_{12} = \frac{3a}{8\ell}\zeta_w R_2\dot{\phi}_2 [ & (2A(\beta) +B(\beta))\cos(\phi_1-\phi_2) \nonumber \\ & + B(\beta)\cos(\phi_1+\phi_2)], \\
\hat{e}_{R1} \cdot \bm{v}_{12} = \frac{3a}{8\ell}\zeta_w R_2\dot{\phi}_2 [ & (2A(\beta) +B(\beta))\sin(\phi_1-\phi_2) \nonumber \\ & + B(\beta)\sin(\phi_1+\phi_2) ].
\end{align}
The rest of the calculation can be carried out in exactly the same way as in \cite{Niedermayer:2008fk} and the final result is
\begin{align}
\dot{\phi}_1 &= \omega_1 -\rho\omega_2J(\phi_1,\phi_2;\beta)-\rho\alpha\omega_1\omega_2K(\phi_1,\phi_2;\beta), \\
\dot{\phi}_2 &= \omega_2 -\rho\omega_1J(\phi_2,\phi_1;\beta)-\rho\alpha\omega_1\omega_2K(\phi_2,\phi_1;\beta), 
\end{align}
where now $\rho = 3a\zeta_w/8\ell$, $\alpha = \bar{\omega}\zeta_0\zeta_w/\lambda$, and 
\begin{align}
J(\phi_i,\phi_j:\beta) &= - [ (2A(\beta)+ B(\beta))\cos(\phi_i-\phi_j) \nonumber \\ & + B(\beta)\cos(\phi_i+\phi_j) ] \\
K(\phi_i,\phi_j:\beta) &= ( 2A(\beta)+B(\beta))\sin(\phi_i-\phi_j) \nonumber \\ & + B(\beta)\sin(\phi_i+\phi_j).
\end{align}
For example, this means that the phase difference $\chi = \phi_2-\phi_1$ and phase sum $\Phi = \phi_1+\phi_2$ evolve according to
\begin{align}
\dot{\chi} = (& \omega_2-\omega_1)[1+\rho J(\phi_1,\phi_2;\beta)]  \nonumber \\ & - 2\rho \alpha \omega_1\omega_2(2A(\beta)+B(\beta))\sin(\chi), \\
\dot{\Phi} = (& \omega_1+\omega_2) [ 1+\rho(2A(\beta)+B(\beta))\cos\chi \nonumber \\ & +\rho B(\beta)\cos\Phi ] - 2\rho\alpha\omega_1\omega_2B(\beta)\sin\Phi.
\end{align}
To the first order in the small quantities $\Delta\omega = \omega_2-\omega_1$ and $\rho$, and averaging over a ``natural'' timescale of the fast variable $\Phi$ we get
\begin{align}
\dot{\chi} &= \Delta\omega - 2\alpha\omega_1\omega_2\,\rho(2A(\beta)+B(\beta))\sin \chi, \label{phase_drift} \\ 
\langle \dot{\Phi} \rangle &= (\omega_1+\omega_2)\left[1+\rho(2A(\beta)+B(\beta))\cos \chi \right]. 
\end{align}
These functions can be rewritten as 
\begin{align}
\dot{\chi} &= \Delta \omega + \tilde{D}(\chi), \\
\langle{\dot{\Phi}}\rangle &= (\omega_1 + \omega_2) + \tilde{S}(\chi),
\end{align}
where $\tilde{D}(\chi) = D_0 \sin \chi$ and $\tilde{S}(\chi) = S_0 \cos \chi$. As in \cite{Niedermayer:2008fk}, time has been rescaled according to the mean angular speed $\bar{\omega}$, and both $\omega_i$ are measured in units of $\bar{\omega}$. Rewriting Eq.~\eqref{phase_drift} in dimensional units yields
\begin{equation}
\dot{\chi} = \Delta \omega - \frac{3a}{4 \ell} \frac{\zeta_0 \zeta_w^2}{\lambda} \omega_1 \omega_2 \big [ 2A(\beta)+B(\beta) \big] \sin(\chi). \label{phase_drift_dimensional}
\end{equation}
This is of the form $\dot{\chi} = \Delta \omega - C \sin (\chi)$, with $C>0$. If $|\Delta \omega| < C$ then a stable fixed point $\dot{\chi}=0$ exists. Conversely, for $|\Delta \omega|>C$, a cycle-averaged phase drift will occur. We use the following relation
\begin{equation}
\int_0^{2 \pi} \big[ a - b \sin \chi \big]^{-1} d \chi = \frac{2 \pi}{\sqrt{a^2-b^2}}, \quad \text{for} \ |a| > |b|
\end{equation}
to find the time-averaged phase drift
\begin{equation}
\dot{\chi}_{\text{av}} = \sqrt{ (\Delta \omega)^2 - \bigg(\frac{3a}{4l} \frac{\zeta_0 \zeta_w^2}{\lambda} \omega_1 \omega_2 \big[ 2A(\beta) + B(\beta) \big] \bigg)^2}.
\end{equation}
For each rotor (now indexed by $i \in \{ 0,1\}$), the intrinsic angular frequency is given by $\omega_i = F_i / (\zeta_0 \zeta_w R_0)$ and so the above equations reads: 
\begin{equation}
\dot{\chi}_{\text{av}} = \sqrt{ \bigg( \frac{F_1-F_0}{R_0 \zeta_0 \zeta_w} \bigg)^2 - \bigg(\frac{3a}{4 \ell} \frac{F_0 F_1}{\lambda \zeta_0 R_0^2} \big[ 2A(\beta) + B(\beta) \big] \bigg)^2}. \label{phase_drift_S1}
\end{equation}
Since a detuning factor $D$ is included so that the driving force is $F_i = F_{\text{dr}} D^{i-1/2}$, the above equation can be written as
\begin{equation}
\dot{\chi}_{\text{av}} = \frac{F_{\text{dr}}}{R_0 \zeta_0} \sqrt{ \frac{(D-1)^2}{D\,\zeta_w^2} - \bigg(\frac{3a}{4 \ell} \frac{F_{\text{dr}}}{\lambda R_0} \big[ 2A(\beta) + B(\beta) \big] \bigg)^2}. \label{phase_drift_S2}
\end{equation}
The threshold value of $D$ beyond which the rotors' phase difference will drift can be calculated explicitly.

\subsection{Varying chain length}

\begin{figure}[htp!]
\begin{center}
\includegraphics[width=\columnwidth]{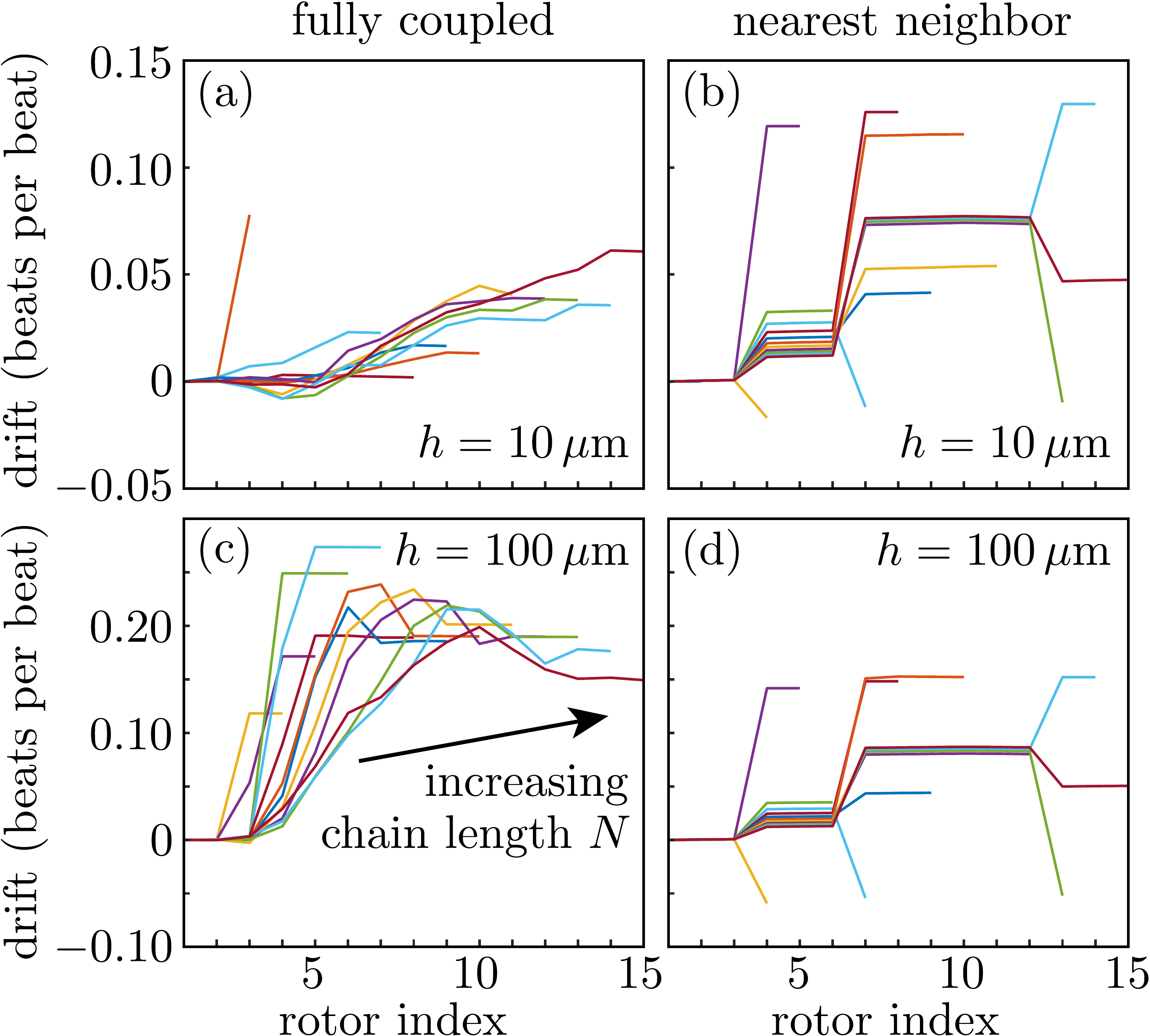} 
\caption{(color online). Average phase drift with respect to the first rotor, for chains of different lengths $N \in \{2,15\}$, at two different heights $h \in \{10\,\mu \text{m}, 100\,\mu \text{m} \}$, and subject to either full hydrodynamic coupling or nearest neighbor interactions only. The end-to-end detuning is fixed at $5\%$ in each case, the radial spring stiffness is $\lambda = 4.5 \, \text{pN}/\mu \text{m},$ and all other parameters are as in Fig.~4.}
\label{figS2}
\end{center}
\end{figure}

In order to assess the generality of the results presented in the main text, we used numerical simulations to explore the effect of changing the number of rotors, $N$, present in the linear array (see Fig.~1a). 
Figure~\ref{figS2} shows the average phase drift (measured in beats per beat) with respect to the first rotor, along chains of different length, $N \in \{2,15\}$. Each chain has a fixed detuning of $5\%$ between the end rotors. For each height $h=10\,\mu \text{m}$ and $h=100\,\mu \text{m}$, simulations were conducted with full hydrodynamic coupling and nearest neighbor coupling only.

The results for $N=6$ are representative of the dynamics across a range of chain lengths. For chains in which rotors are coupled through nearest neighbor interactions, the rotors tend to phase-lock in clusters of 2-5 rotors. As discussed in the main text, the nearest neighbor results are fairly insensitive to changes in $h$, shown here by the similarly between the results of Fig.~\ref{figS2}b and d. In stark contrast, the chains in which rotors are fully coupled to one another through hydrodynamic interactions exhibit qualitatively different behavior at different heights.

\newpage 
\subsection{Truncation of hydrodynamic interactions}

\begin{figure}[htp!]
\begin{center}
\includegraphics[width=\columnwidth]{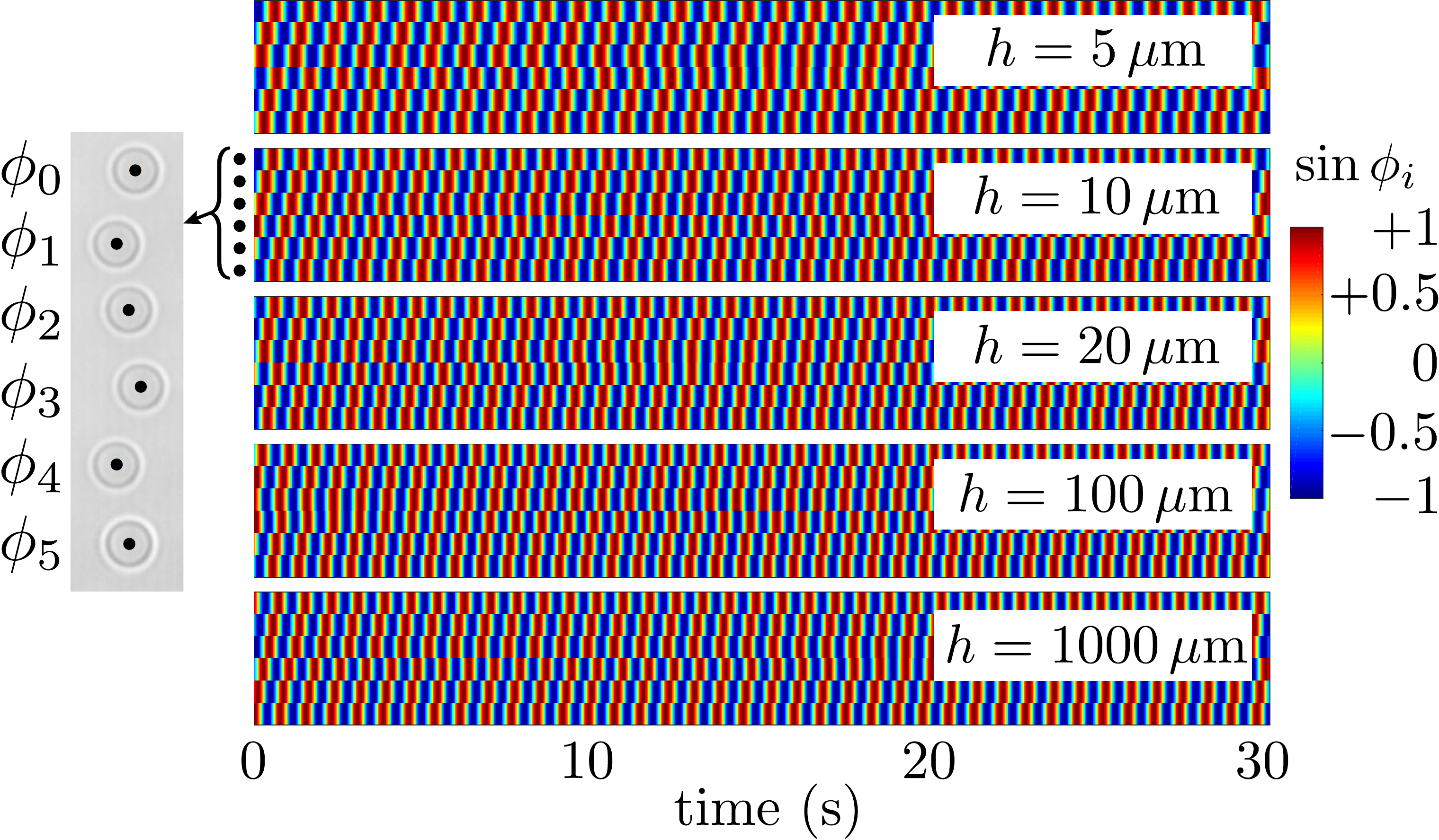} 
\caption{(color online). Kymographs showing the phase $\sin \phi_i$ along the linear chain of model rotors, coupled hydrodynamically through the Blake tensor, but with interactions artificially restricted to be nearest neighbor. The radial spring stiffness is $\lambda = 4.5 \, \text{pN/}\mu\text{m}$ and all other parameters are as in Fig.~4.}
\label{figS3}
\end{center}
\end{figure}

Figure~\ref{figS3} shows the results of deterministic numerical simulations, with hydrodynamic interactions truncated to be nearest neighbor in nature (see also Fig.~4d). 
The dynamics are almost completely insensitive to changes in $h$, across several orders of magnitude.

\begin{thebibliography}{10}

\bibitem{Dorfler2014}
F. D{\"{o}}rfler and F. Bullo, 
{Automatica} {\bf 50}, 1539 (2014).

\bibitem{Mirollo1990}
R.~E. Mirollo and S.~H. Strogatz, 
{SIAM J. Appl. Math.} {\bf 50}, 1645 (1990).

\bibitem{Toiya2010}
M. Toiya, H.~O. Gonz{\'{a}}lez-Ochoa,  V.~K. Vanag, S. Fraden and I.~R. Epstein,
{J. Phys. Chem. Lett.} {\bf 1}, 1241  (2010).

\bibitem{Son2015}
K. Son, D.~R. Brumley, and R. Stocker,  
{Nat. Rev. Micro.} {\bf 13}, 761 (2015).

\bibitem{Goldstein2015}
R.~E. Goldstein, 
{Annu. Rev. Fluid Mech.} {\bf 47}, 343 (2015).

\bibitem{Quaranta2016}
G. Quaranta, M.-E. Aubin-Tam, and D. Tam,
{Phys. Rev. Lett.}{\bf 115}, 238101 (2016).

\bibitem{Sleigh1962}
M.~A. Sleigh,
{\em The Biology of Cilia and Flagella},
(Pergamon Press, Oxford, 1962).

\bibitem{Button:2012}
B. Button, L. Cai, C. Ehre, M. Kesimer, D.~B. Hill, J.~K. Sheehan, R.~C. Boucher,  and M. Rubinstein, 
{Science} {\bf 337}, 937 (2012).

\bibitem{Brumley2014}
D.~R. Brumley, K.~Y. Wan, M. Polin, and R.~E. Goldstein,  
{eLife} {\bf 3}, e02750 (2014).

\bibitem{Knight_Jones:1954}
E.~W. Knight-Jones, 
{Quart. J. Micro. Sci.} {\bf 95}, 503 (1954).

\bibitem{Brumley:2012}
D.~R. Brumley, M. Polin, T.~J. Pedley,  and R.~E. Goldstein,  
{Phys. Rev. Lett.} {\bf 109}, 268102 (2012).

\bibitem{Brumley:2015_JRSI}
D.~R. Brumley, M. Polin, T.~J. Pedley, and R.~E. Goldstein, 
{J. R. Soc. Interface} {\bf 12}, 20141358 (2015).

\bibitem{Niedermayer:2008fk}
T. Niedermayer, B. Eckhardt, and P. Lenz, 
{Chaos} {\bf 18}, 37128 (2008).

\bibitem{Vilfan:2006uq}
A. Vilfan and F. J\"{u}licher, 
{Phys. Rev. Lett.} {\bf 96}, 058102  (2006).

\bibitem{Uchida:2011kx}
N. Uchida and R. Golestanian 
{Phys. Rev. Lett.} {\bf 106}, 058104 (2011).

\bibitem{Wollin:2011}
C. Wollin and H. Stark, 
{Eur. Phys. J. E} {\bf 34}, 42 (2011).

\bibitem{Gueron:1999}
S. Gueron and K. Levit-Gurevich,
{Proc. Natl. Acad. Sci. USA} {\bf 96}, 12240 (1999).

\bibitem{Lagomarsino03}
M. Cosentino~Lagomarsino, P. Jona, and B. Bassetti, 
{Phys. Rev. E} {\bf 68}, 021908 (2003).

\bibitem{Osterman:2011}
N. Osterman and A. Vilfan, 
{Proc. Natl. Acad. Sci. USA} {\bf 108},
  15727 (2011).

\bibitem{Elgeti:2013}
J. Elgeti and G. Gompper, 
{Proc. Natl. Acad. Sci. USA} {\bf 110},
  4470 (2013).

\bibitem{Bruot13}
N. Bruot, and P. Cicuta, 
{J. R. Soc. Interface} {\bf 10}, 20130571 (2013).

\bibitem{Kavre2015}
I. Kavre, A. Vilfan, and D. Babi{\v{c}}, 
{Phys. Rev. E} {\bf 91}, 031002 (2015).

\bibitem{Bruot2016}
N. Bruot and P. Cicuta, 
{Annu. Rev. Condens. Matt. Phys.} {\bf 7}, 323 (2016).

\bibitem{Abrams2004}
D.~M. Abrams and S.~H. Strogatz,  
{Phys. Rev. Lett.} {\bf 93}, 174102 (2004).

\bibitem{Martens2013}
E.~A. Martens, S. Thutupalli, A. Fourri{\`{e}}re, and O. Hallatschek,  
{Proc. Natl. Acad. Sci. USA} {\bf 110},
  10563 (2013).

\bibitem{Leoni2009}
M. Leoni, J. Kotar, B. Bassetti, P. Cicuta, and M.~C. Lagomarsino, 
{Soft Matter} {\bf 5}, 472  (2009).

\bibitem{Kotar2013}
J. Kotar, L. Debono, N. Bruot, S. Box, D. Phillips, D. Simpson, S. Hanna, and P. Cicuta, 
{Phys. Rev. Lett.} {\bf 111}, 228103 (2013).

\bibitem{SM}
Supplementary Material.

\bibitem{Happel_Brenner}
J. Happel and H. Brenner, 
{\em Low Reynolds Number Hydrodynamics}
(Kluwer, Dordrecht, 1991).

\bibitem{Ahnert2008}
K. Ahnert and A. Pikovsky, 
{Chaos} {\bf 18}, 37118 (2008).

\bibitem{Pikovsky:2003}
A. Pikovsky, M. Rosenblum, and J. Kurths, {\em Synchronization: A Universal Concept in Nonlinear Sciences} 
(Cambridge University Press, Cambridge, 2003).

\bibitem{Blake:1971uq}
Blake, J.~R. (1971)
{\em Mathematical Proceedings of the Cambridge Philosophical Society} {\bf 70(2)}, 303--310.

\end{thebibliography}
\end{document}